# Advanced magnetocaloric microwires: What does the future hold?

**Hongxian Shen[1, 2], Nguyen Thi My Duc[1, 3], Hillary Belliveau[1], Lin Luo[2], Yunfei Wang[4], Jianfei Sun[2], Faxiang Qin[4], Manh-Huong Phan[1*]**

[1]*Department of Physics, University of South Florida, Tampa, FL, 33620, USA*

[2]*School of Materials Science and Engineering, Harbin Institute of Technology, Harbin 150001, P. R. China*

[3]*The University of Danang, University of Science and Education, 459 Ton Duc Thang Street, Lien Chieu County, Danang City, Vietnam*

[4]*Institute for Composites Science Innovation (InCSI), School of Materials Science and Engineering, Zhejiang University, 38 Zheda Road, Hangzhou, 310027, P. R. China*

**Magnetic refrigeration (MR) based on the magnetocaloric effect (MCE) is a promising alternative to conventional vapor compression refrigeration techniques. The cooling efficiency of a magnetic refrigerator depends on its refrigeration capacity and operation frequency. Existing refrigerators possess limited cooling efficiency due to the low operating frequency (around tens of Hz). Theory predicts that reducing geometrical effects can increase the operation frequency by reducing the relaxation time of a magnetic material. As compared to other shapes, magnetocaloric wires transfer heat most effectively to a surrounding environment, due to their enhanced surface area. The wire shape also yields a good mechanical response, reducing the relaxation time and consequently increasing the operation frequency of the cooling device. Experiments have validated the theoretical predictions. By assembling microwires with different magnetocaloric properties and Curie temperatures into a laminate structure, a table-like magnetocaloric bed can be created and**

[*]*Corresponding author: Email: phanm@usf.edu*

**used as an active cooling device for micro-electro-mechanical system (MEMS) and nano-electro-mechanical system (NEMS). This paper assesses recent progress in the development of magnetocaloric microwires and sheds light on the important factors affecting the magnetocaloric behavior and cooling efficiency in microwire systems. Challenges, opportunities, and strategies regarding the development of advanced magnetocaloric microwires are also discussed.**

***Keywords:*** Magnetic microwires, Magnetocaloric effect, Magnetic refrigeration, Refrigerant capacity.

## 1. Magnetic refrigeration based on the magnetocaloric effect

The MCE is a phenomenon where a magnetic material exhibits a change in temperature or magnetic entropy when subjected to an external magnetic field. MR technology based on the MCE is energy-saving, highly cooling efficient, and environmentally friendly, which makes it a promising platform to replace gas compression (GC) refrigeration techniques [1-6]. Unlike GC-based refrigerators, MR does not use harmful gases (e.g., hydrofluorocarbons) that can jeopardize the ozonosphere and gets rid of complicated GC device components, making a simple and compacted MR device with low vibration and noise [1, 4]. Therefore, it is considered a future green solid-state cooling technology [6].

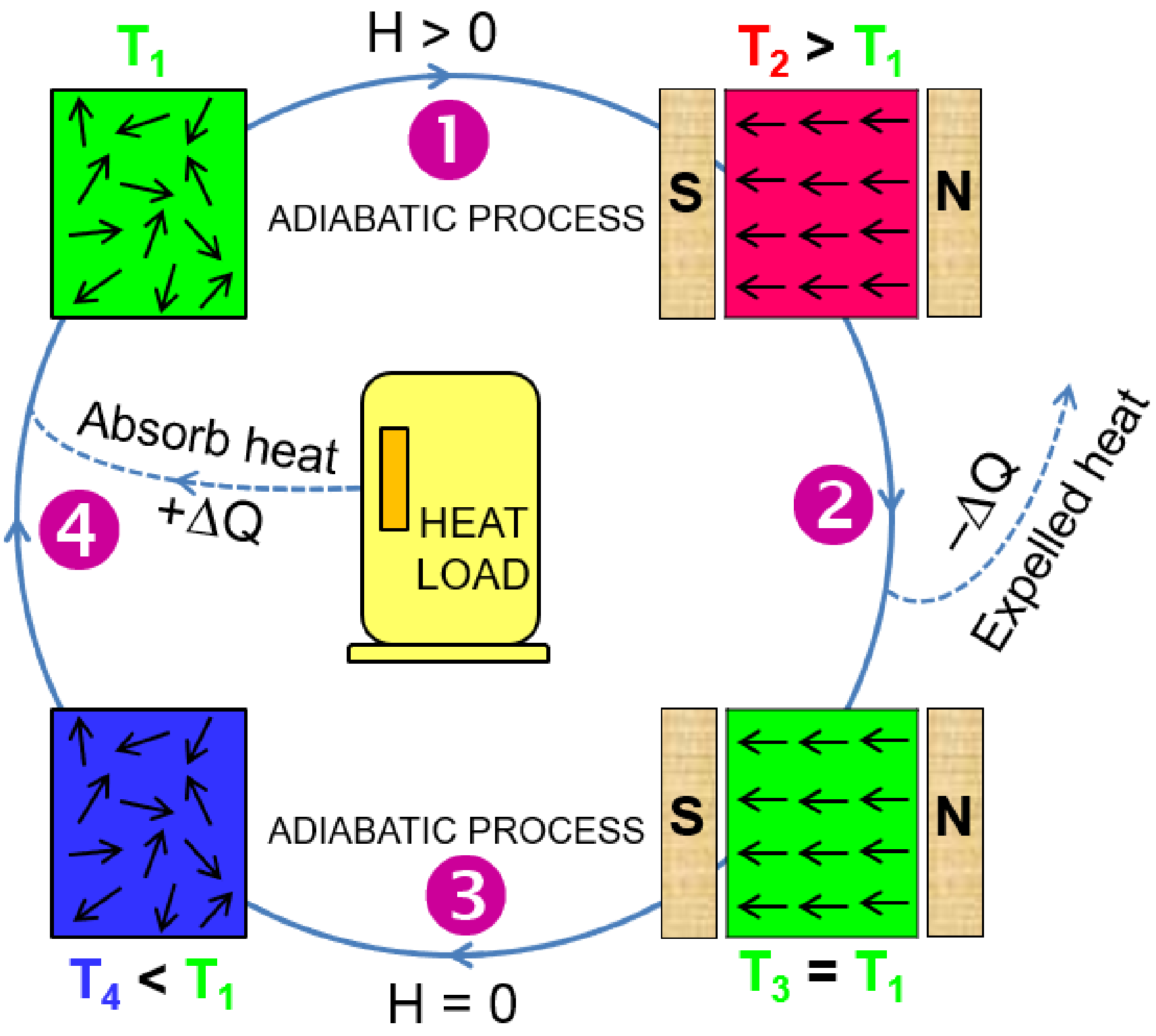

**Fig. 1. The four stages of a MR cycle: (1) adiabatic magnetization, (2) heat removal, (3) adiabatic demagnetization, and (4) cool refrigerator contents.**

Figure 1 illustrates a full cooling cycle principle of a magnetic refrigerator based on the MCE [7]. In the absence of an applied magnetic field, the magnetic material has a temperature $T_1$. As the material is adiabatically magnetized, the magnetic moments of the ferromagnetic material orient themselves in the direction of the applied field. The alignment of the magnetic moments causes a decrease in the spin entropy and therefore the lattice entropy is increased. This increase in lattice entropy heats the material; the temperature of the material increases from $T_1$ to $T_2$=$T_1$+$\Delta T$. This excess heat ($\Delta T$) is then removed by heat exchange with a cooler body, thus reducing the temperature of the material to $T_3$=$T_1$. The removal of the applied field causes the magnetic moment entropy to increase and the lattice entropy and hence the material temperature to decrease to $T_4$=$T_1$-$\Delta T$. The increase/decrease in lattice entropy due to the decrease/increase in magnetic entropy alters temperature of the material and this can be used to refrigerate. It is worth noting that the heat transfer during the MCE process is dependent upon the entropy change in the magnetic refrigerant element [1, 4, 5]. Therefore, a material system exhibiting a large magnetic entropy change ($\Delta S_M$) is highly valuable to MR technology [6].

## 2. Whate make magnetocaloric microwires promising for MR?

The geometric effect of refrigerants used in magnetic coolers has been overlooked by researchers looking to improve MR technology [1, 7-10]. Solid magnetocaloric materials can be fabricated in different forms: powder particles, thin films, or wires [1, 4, 5]. To improve magnetic cooling, it's very important to design them in wire form so that they can release heat better to surroundings because of their enhanced surface areas [9, 10].

M.D. Kuzmin (2007) predicted that by reducing the dimensions of a magnetic refrigerant the cooling power of the device could be increased by increasing the operating frequency [9]. His theoretical study has shown that shaping magnetic refrigerants in the form of spherical or irregular particles is inefficient, due to their high losses on viscous resistance and demagnetization and suggests that the wire geometry could be utilized instead. Mechanical instability of the refrigerant can result in a significant loss of heat throughout due to unequal distribution of flow. In this context, the use of a bundle of magnetocaloric wires (e.g. Gd wires of ~50 μm diameter) has been proposed to be more desirable because this configuration enables higher mechanical stability and lower porosity while simultaneously increasing the operating frequency of the magnetic refrigerant material [9, 10].

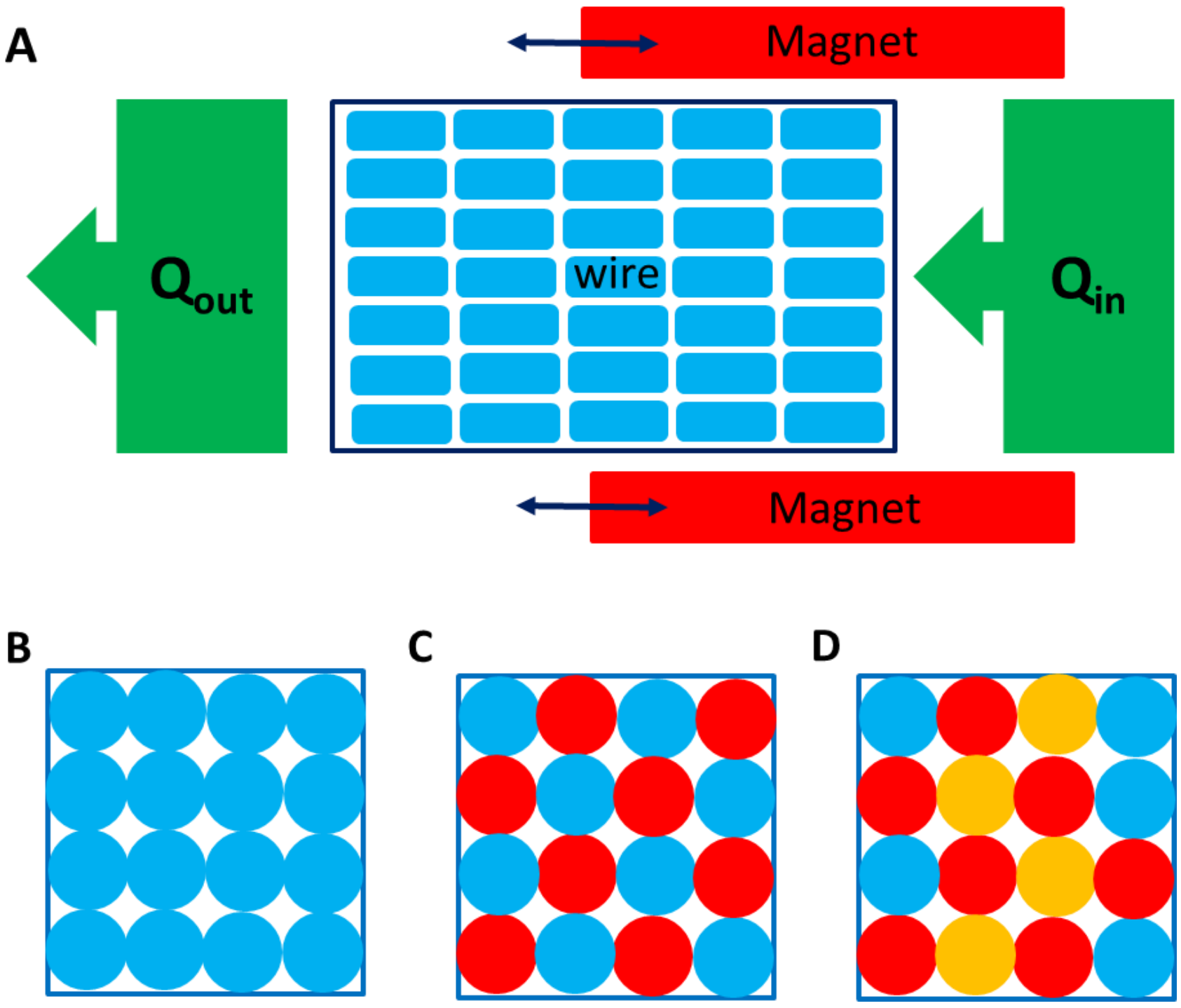

**Fig. 2. (A) Illustration of a simplified active cooling system using magnetocaloric wires, similar to that proposed by D. Vuarnoz and T. Kawanami (2012) [10]. Proposed magnetic beds composed of magnetocaloric microwires with same (B) and different (C, D) compositions (illustrated with different colors) possessing appropriate magnetic entropy changes and Curie temperatures, resulting in enhanced cooling efficiency.**

D. Vuarnoz and T. Kawanami (2012) [10] have theoretically predicted that a magnetic bed made of an array of Gd wires (Fig. 2A) produces a greater temperature difference between its ends, which results in a higher cooling load at a higher cooling efficiency, as compared to a magnetic bed made of Gd particles [10]. Unlike their bulk counterparts, the use of the wires with increased surface areas also allows for a higher heat transfer between the magnetic refrigerant and surrounding liquid [8]. These theoretical studies have opened new areas for MR devices and material design [9, 10]. Recent experiments on various magnetocaloric microwires have confirmed these predictions [11-42]. Since these magnetocaloric microwires can be easily assembled as laminate structures (Fig. 2B-D), they have potential applications as cooling devices for MEMS and NEMS. Instead of using the same type of magnetocaloric microwires (Fig. 2B) to design a magnetic bed, magnetocaloric microwires with different $\Delta S_M$ and Curie temperature values (Fig. 2C, D) can also be assembled to design a table-like magnetocaloric bed with an enhanced *RC* resulting from the broadened $\Delta S_M(T)$ curve.

**3. Criteria for selecting magnetocaloric materials**

The change in magnetic entropy ($\Delta S_M(T,\mu_0 H)$) of a magnetic material with respect to $\mu_0 H$ and $T$ is often considered for MCE studies. The $\Delta S_M(T,\mu_0 H)$ is related to the change in magnetization ($M$) with respect to $T$ through Maxwell relation [4, 5]:

$$\left(\frac{\partial S(T,H)}{\partial H}\right)_T = -\left(\frac{\partial M(T,H)}{\partial T}\right)_H \tag{1}$$

The $\Delta S_M$ caused by varying external magnetic field from $H = 0$ to $H = H_0$ is evaluated as

$$\Delta S_M(T,H_0) = S_M(T,H_0) - S_M(T,0) = \mu_0 \int_0^{H_0} \left(\frac{\partial M(T,H)}{\partial T}\right)_H dH. \tag{2}$$

The adiabatic temperature change, $\Delta T_{ad}$, at a given temperature $T_0$ can be approximately calculated by

$$\Delta T_{ad}(T_0, H_0) \cong -\Delta S_M(T_0, H_0)\frac{T_0}{C(T_0, H_0)} \tag{3}$$

In these above equations, $\mu_0$ is the permeability of vacuum and $(\partial M/\partial T)$ is the derivative of the magnetization with respect to temperature in a constant magnetic field. According to Eq. (2), the magnitude of $\Delta S_M$ depends on both the magnitude of $M$ and $(\partial M/\partial T)_H$. As these values increase, so does the magnitude of $\Delta S_M$, which is desired in the design of magnetic cooling devices. To obtain a large $\Delta S_M$ it is best to use a material with a steep change in $M$ with respect to temperature at the transition temperature because $(\partial M/\partial T)_H$ is closely related to the magnetic order transition. From Eq. (3) a large $\Delta T_{ad}$ can be obtained by having a large $\Delta S_M$ around the magnetic ordering temperature and a small *C(T,H)*. However *C(T,H)* varies widely between materials and as a result the larger $\Delta S_M$ does not necessarily lead to a larger $\Delta T_{ad}$. Therefore, the parameter that characterizes the practicality of a magnetic refrigerant cannot solely be determined by the magnitude of $\Delta S_M$.

To properly compare magnetic refrigerants, it is important to consider the temperature response of $\Delta S_M$ of the refrigerant, namely the refrigerant capacity (*RC*). The *RC* of a magnetic refrigerant material is calculated using [4]:

$$RC = \int_{\mathrm{T_{hot}}}^{\mathrm{T_{cold}}} -\Delta S_{\mathrm{M}}(\mathrm{T})\mathrm{dT}, \qquad (4)$$

The *RC* indicates how much heat can be transferred from the cold end (at $T_{\mathrm{cold}}$) to the hot end (at $T_{\mathrm{hot}}$) of a refrigerator in an ideal thermodynamic cycle. The relative cooling power (RCP), representing an amount of heat transfer between the hot and cold sides in an ideal refrigeration cycle, can be defined as

$$RCP = -\Delta S_{\mathrm{M}}^{\mathrm{max}}\ \delta T_{\mathrm{FWHM}}, \qquad (5)$$

where $\delta T_{\mathrm{FWHM}} = T_{\mathrm{hot}} - T_{\mathrm{cold}}$ is the temperature difference at the full width at half maximum of the magnetic entropy change curve. Both *RC* and *RCP* have been used to assess the usefulness of magnetocaloric materials for MR [1, 4, 5].

It is generally accepted that while first-order magnetic transition materials (FOMT) possess larger values of $\Delta S_M$, they are characteristically limited to narrower temperature ranges [1,4,5]. This is the limiting factor of the operating temperature of this type of magnetic material. However, second-order magnetic transition materials (SOMT) have lower values of $\Delta S_M$, which extends through a broader temperature range. Because of the broader operating temperature range as well as negligible thermal hysteretic losses, the *RC* of SOMT materials is generally larger than that of FOMT materials [4, 5]. Fig. 3 compares $\Delta S_M$ and *RC* values of the same material $Pr_{0.5}Sr_{0.5}MnO_3$ that undergoes the SOMT around the $T_C$ (~250 K) and the FOMT around the $T_{CO}$ (~150 K) [43]. It can be seen that although the $\Delta S_M$ at the FOMT (~6 J/kg K) is three times greater than the $\Delta S_M$ at the SOMT (~2 J/kgK), the RC of the latter (~150 J/kg) is larger than the RC of the former (~120 J/kg), when the magnetic hysteresis loss is subtracted from the RC calculation according to Eq. (4).

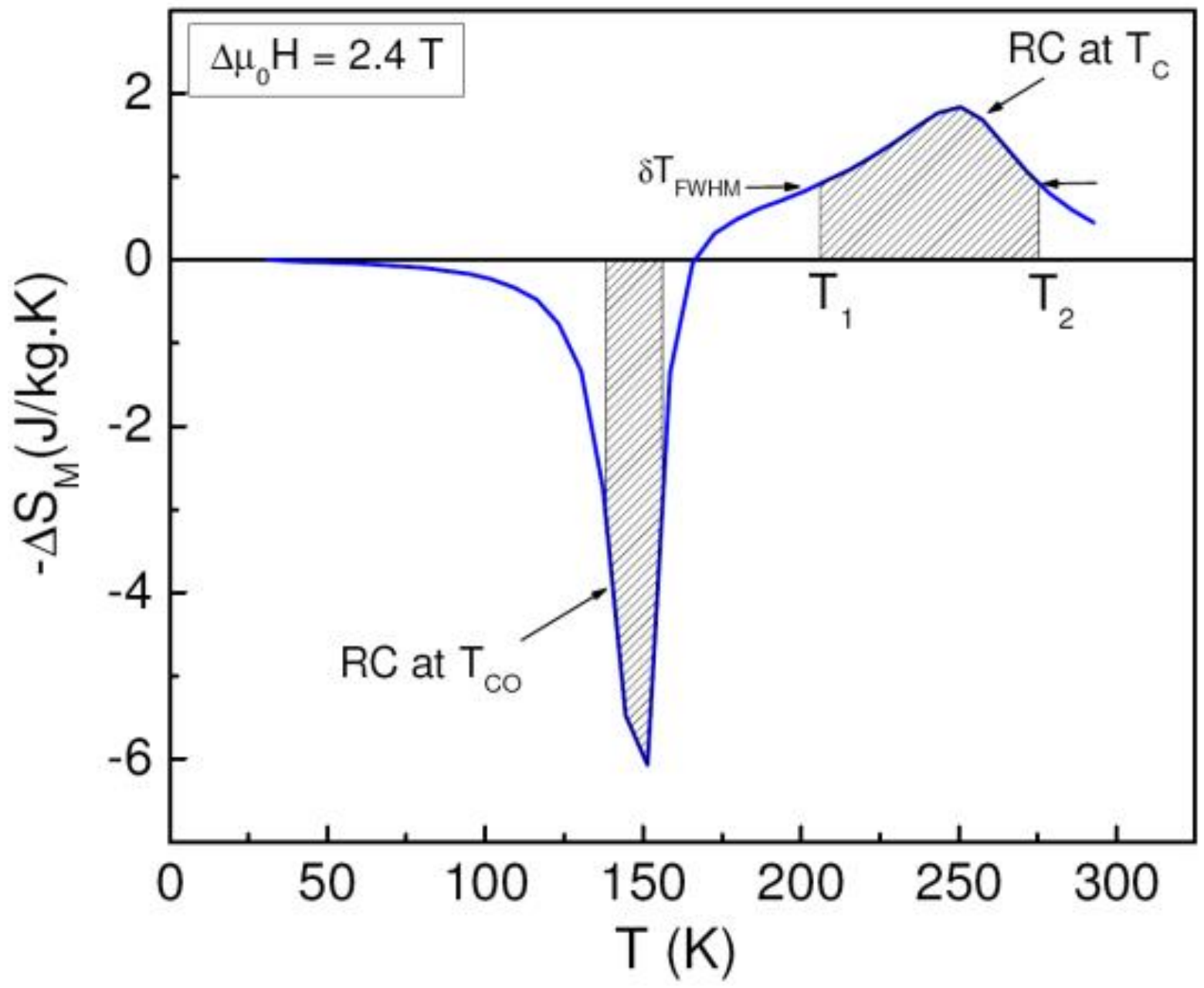

**Fig. 3. The method for calculating the refrigerant capacity (*RC*) from the -$\Delta S_M(T)$ curve using Eq. (4) for the cases of the second-order magnetic transition (SOMT) around the $T_C$ and the first-order magnetic transition (FOMT) around the $T_{CO}$ for the $Pr_{0.5}Sr_{0.5}MnO_3$ compound. Adapted from Ref. [43].**

Among the reported magnetocaloric materials, Gadolinium (Gd) is considered the benchmark material, which shows the best magnetocaloric characteristics for sub-room temperature magnetic cooling applications [1, 4, 6]. From a practical application perspective, however, Gd is too expensive, and it is therefore essential to reduce the amount of Gd used in the MR devices by doping or alloying, such as $Gd_5(Si_xGe_{1-x})_4$ compounds [4]. Such doping or alloying also allows for the tailoring of the structural, magnetic, and magnetocaloric properties of Gd-based materials for MR at different temperature regimes [1, 4]. The majority of the research in this field is therefore focused on the development of Gd-based microwires for MR in the nitrogen temperature regime [11-29]. Other types of magnetocaloric microwires such as Heusler and high-entropy alloy microwires have recently been designed and fabricated, and some of them exhibit excellent magnetocaloric properties in the room temperature region [30-42]. We assess below (section 5.1) the magnetocaloric properties of these microwires and their usefulness for MR applications.

**4. Effects of reduced dimensionality on the magnetocaloric behavior and cooling efficiency**

A clear understanding of the effects of reduced dimensionality on the magnetocaloric response of a magnetic material is essential to optimizing the performance of a cooling device [8, 44]. To gain insights into these effects, Fig. 4 compares the magnetic and magnetocaloric properties of $Gd_{53}Al_{24}Co_{20}Zr_3$ in various forms of bulk alloys, a single microwire, and a multiwire array. It can be seen that the shape of the *M-T* curves for the single-wire and multiwire samples is rather flat, which is in sharp contrast with the sharp transition at the $T_C$ for the bulk sample (Fig. 4A). However, this feature is only observed when the applied magnetic field is small (less than 500 Oe). As compared to the single wire, the *M-T* curve becomes broader near the $T_C$ in the multiwire array, which can arise from the enhanced magnetostatic interaction between wires in the multi-wire system.

The Curie temperature values of the samples are determined by the d*M*/d*T* vs. *T* curves, which show a slight reduction in $T_C$ in the Gd-based materials with reduced dimensions. Interestingly, the *M-H* data (Fig. 4B) shows that the magnetization of the single wire is more dominant at low field range (<0.5 T or 5000 Oe) compared to its bulk counterpart, while an opposite trend is observed at higher magnetic fields. The multiwire system shows the largest magnetization among these samples (Table 1). This can be attributed to the enhanced multiwire interaction, where spins near the surfaces of the wires align, giving rise to an enhancement in the magnetization of the system. Owing to the reduced dimensionality and shape anisotropy effects, the single and multi-wire samples show a faster approach to saturation of magnetization with respect to an applied magnetic field, as compared to their bulk counterpart. As a result, the multi-wire sample shows the largest $\Delta S_M$ and RC values among the samples investigated (Fig. 4C, D), while it is found to be almost equal for the bulk and single-wire samples (Table 1).

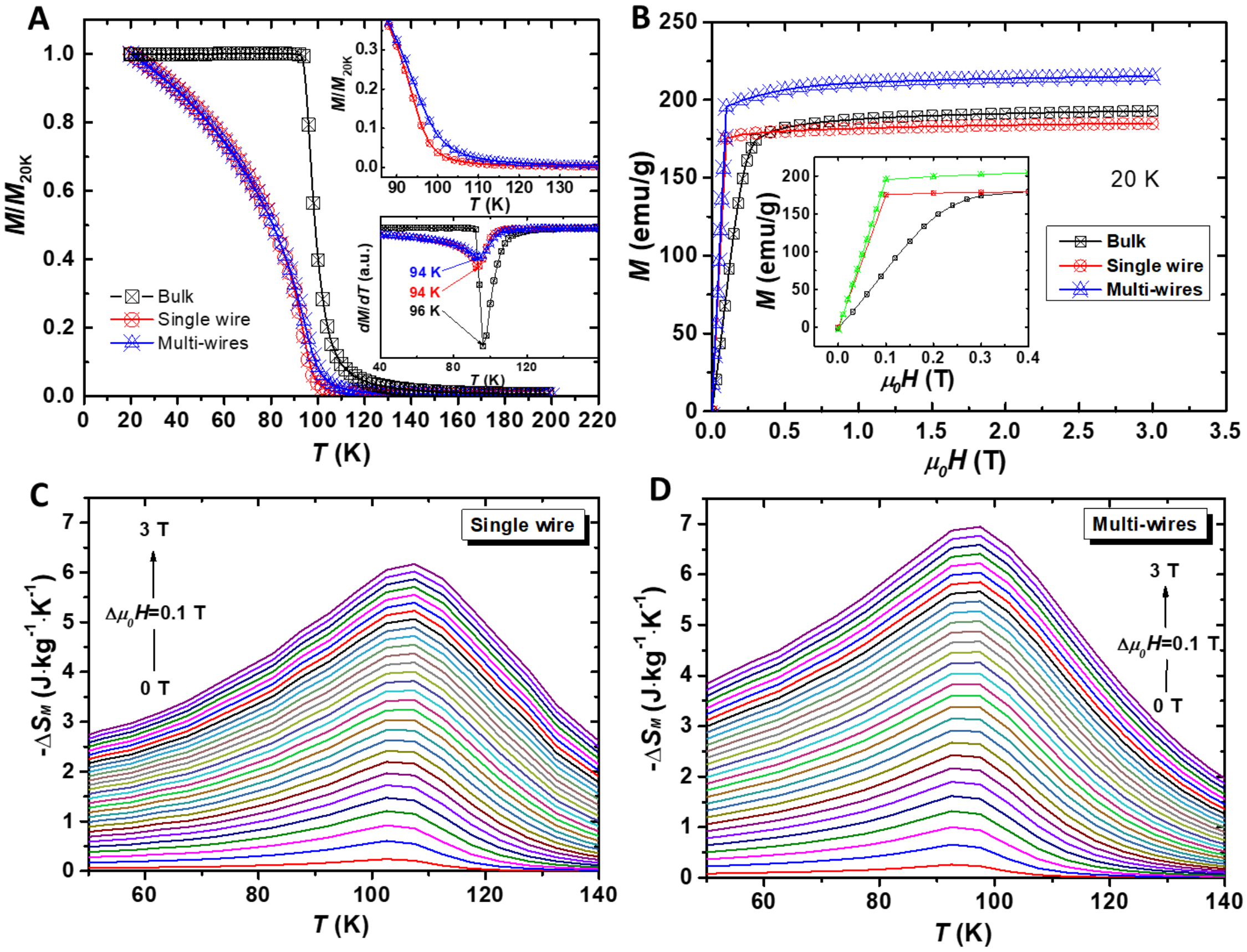

**Fig. 4. (A) Temperature and (B) magnetic field dependence of the magnetization for $Gd_{53}Al_{24}Co_{20}Zr_3$ functionalized as bulk alloys, a single microwire, and a multiwire array. Temperature dependence of the magnetic entropy change -$\Delta S_M(T)$ around the $T_C$ for (C) the single wire and (D) the multiwire system.**

These findings are important, highlighting the superior properties of the magnetocaloric microwire arrays that can be used as magnetic beds in active magnetic refrigerators. Nonetheless, it is worth noting that the heterogeneity of wire diameters affects wire-wire interaction and therefore MCE behavior. If this can be well controlled, the heterogeneity effect is minimal, and the resulting $\Delta S_M$ and *RC* can be optimized in the multiwire array. As compared to the single wire, the multiwire array shows a significantly enhanced $\Delta S_M$ and *RC*. This feature has been mostly observed for the case of Gd-based microwires, which appears to be opposite for the cases of MnFePSi and Heusler alloy systems, due to composition inhomogeneity. Therefore, it is important to provide specific information about wire samples when reporting MCE results in journals. Also, it remains unclear how $\Delta S_M$ and *RC* vary with the wire diameter and/or the number of wires in an array. The case might become more complicated if different types of wires are used in the array. These important issues need to be addressed in future studies.

In the following sections, unless specified, we have assessed the magnetocaloric properties of different magnetic multiwire systems. It should be noted herewith that the comparison of $\Delta S_M$ and RC values between reported multi-wire systems is rather relative, since the total number of wires used in each multiwire array can vary.

**Table 1. Maximum entropy change, $\left|\Delta S_M^{max}\right|$, Curie temperature, $T_C$, refrigerant capacity (*RC*), and relative cooling power (*RCP*) for the as-prepared amorphous and annealed wire samples. Values from microwires of other compositions, bulk glasses, Gd and $Gd_5Si_2Ge_{1.9}Fe_{0.1}$ are included for comparison.**

| Microwires | $T_C$ (K) | $\mu_0\Delta H$ (T) | $\left|\Delta S_M^{max}\right|$ (J/kg K) | *RC* (J/kg) | *RCP* (J/kg) | Ref. |
|---|---|---|---|---|---|---|
| $Gd_{53}Al_{24}Co_{20}Zr_3$ (SW, Amorphous) | 94 | 5 | 8.8 | 600 | 774 | [16] |
| | | 2 | 4.3 | 220 | 296 | |
| $Gd_{53}Al_{24}Co_{20}Zr_3$ (SW, at 100°C) | 94 | 5 | 9.5 | 687 | 893 | [16] |
| | | 2 | 4.7 | 285 | 348 | |
| $Gd_{53}Al_{24}Co_{20}Zr_3$ (SW, at 200$^o$C) | 93 | 5 | 8.0 | 629 | 744 | [16] |
| | | 2 | 3.8 | 243 | 307 | |
| $Gd_{53}Al_{24}Co_{20}Zr_3$ (SW, at 300$^o$C) | 92 | 5 | 5.1 | 396 | 525 | [16] |
| | | 2 | 2.4 | 144 | 184 | |
| $Gd_{53}Al_{24}Co_{20}Zr_3$ (B, Amorphous) | 95 | 5 | 9.6 | 690 | - | [11] |
| | 95 | 3 | 6.2 | 340 | - | |
| $Gd_{53}Al_{24}Co_{20}Zr_3$ (MW, Amorphous) | 94 | 5 | 10.3 | 733 | - | [11] |
| | 94 | 3 | 6.9 | 420 | - | |
| $Gd_{50}Co_{20}Al_{30}$ (MW, A+NC) | 86 | 5 | 10.1 | 672 | 896 | [15] |
| $Gd_{55}Co_{20}Al_{25}$ (MW, A+NC) | 100 | 5 | 10.1 | 653 | 870 | [15] |
| $Gd_{60}Al_{20}Co_{20}$ (MW, A+NC) | 109 | 5 | 10.1 | 681 | 908 | [15] |
| $Gd_{55}Al_{20}Co_{25}$ (MW, Amorphous) | 110 | 5 | 9.69 | 580 | 804 | [44] |
| $Gd_{60}Co_{15}Al_{25}$ (MW, Amorphous) | 100 | 5 | 9.73 | 732 | 976 | [45] |
| $Gd_{60}Co_{25}Al_{15}$ (R, Amorphous) | 125 | 5 | 10.1 | 645 | 860 | [46] |
| $Gd_{60}Fe_{20}Al_{20}$ (MW, Amorphous) | 202 | 5 | 4.8 | 687 | 900 | [20] |
| $Gd_{65}Fe_{20}Al_{15}$ (R, Amorphous) | 182 | 5 | 5.8 | 545 | 726 | [47] |
| $Gd_{55}Fe_{15}Al_{30}$ (R, Amorphous) | 158 | 5 | 5.01 | 555 | 741 | [47] |
| $Gd_{55}Fe_{20}Al_{25}$ (R, Amorphous) | 190 | 5 | 4.67 | 651 | 868 | [47] |
| $Gd_{55}Fe_{25}Al_{20}$ (R, Amorphous) | 230 | 5 | 3.77 | 608 | 811 | [47] |
| $Gd_{95}Fe_{2.8}Al_{2.2}$ (GR+NC) | 232 | 5 | 4 | 551 | - | [48] |
| $Gd_{55}Co_{30}Al_{15}$ | 127 | 5 | 9.71 | 573 | 702 | [23] |
| $Gd_{55}Co_{30}Ni_{5}Al_{10}$ | 140 | 5 | 8.91 | 532 | 668 | [23] |
| $Gd_{55}Co_{30}Ni_{10}Al_{5}$ | 158 | 5 | 7.68 | 523 | 653 | [23] |
| $Gd_{73.5}Si_{13}B_{13.5}/GdB_{6}$ (MW, A+NC) | 106 | 5 | 6.4 | 790 | 885 | [21] |
| $Gd_3Ni/Gd_{65}Ni_{35}$ (MW, A+NC) | 120 | 5 | 9.64 | 742 | - | [22] |
| $Gd_{50}$-$(Co_{69.25}Fe_{4.25}Si_{13}B_{13.5})_{50}$ | 170 | 5 | 6.56 | 625 | 826 | [19] |
| $Gd_{59.4}Al_{19.8}Co_{19.8}Fe_1$ (MW, Amorphous) | 113 | 5 | 10.33 | 748 | 1006 | [14] |
| $Mn_{1.26}Fe_{0.60}P_{0.48}Si_{0.52}$ (MW, PC) | 141 | 5 | 4.64 | - | - | [35] |
| $Ni_{48}Mn_{26}Ga_{19.5}Fe_{6.5}$ (MW, PC) | 361 | 5 | 4.7 | - | - | [30] |
| $Ni_{49.4}Mn_{26.1}Ga_{20.8}Cu_{3.7}$ (MW, PC) | 359 | 5 | 8.3 | 78 | - | [31] |
| $Ni_{50.5}Mn_{29.5}Ga_{20}$ | 368 | 5 | 18.5 | 63 | - | [32] |
| $Ni_{50.6}Mn_{28}Ga_{21.4}$ | 340-370 | 5 | 5.2 | 240 | - | [32] |
| $Ni_{45.6}Fe_{3.6}Mn_{38.4}Sn_{12.4}$ (MW, A+NC) | 270 (F)<br>300 (S) | 5<br>5 | 15.2<br>4.3 | 146<br>175 | 182<br>215 | [34] |
| DyHoCo (MW, Amorphous) | 35 | 5 | 11.2 | 417 | 530 | [37] |
| HoErCo (MW, Amorphous) | 16 | 5 | 15 | 527 | 600 | [42] |
| HoErFe (MW, A+NC) | 44 | 5 | 9.5 | 450 | 588 | [36] |
| $Gd_{60}Al_{20}Co_{20}$ (MW, A+NC) | 113 | 5 | 10.12 | 698.28 | 936.19 | [25] |
| $(Gd_{60}Al_{20}Co_{20})_{99}Ni_1$ (MW, A+NC) | 111 | 5 | 10.98 | 725.49 | 970.89 | [25] |
| $(Gd_{60}Al_{20}Co_{20})_{97}Ni_3$ (MW, A+NC) | 109 | 5 | 11.06 | 746.84 | 1000.50 | [25] |
| $(Gd_{60}Al_{20}Co_{20})_{95}Ni_5$ (MW, A+NC) | 109 | 5 | 11.57 | 834.14 | 1138.16 | [25] |
| $(Gd_{60}Al_{20}Co_{20})_{93}Ni_7$ (MW, A+NC) | 108 | 5 | 10.77 | 733.48 | 977.65 | [25] |
| $Gd_{36}Tb_{20}Co_{20}Al_{24}$ (MW, Amorphous) | 91 | 5 | 12.36 | 731 | 948 | [26] |
| $Gd_{36}Tb_{20}Co_{20}Al_{24}$ (MW, A+NC) | 81 | 5 | 8.8 | 500 | 625 | [26] |
| $(Gd_{36}Tb_{20}Co_{20}Al_{24})_{99}Fe_1$ (MW, A+NC) | 94 | 5 | 8.5 | 510 | 635 | [26] |
| $(Gd_{36}Tb_{20}Co_{20}Al_{24})_{98}Fe_{2}$ (MW, A+NC) | 100 | 5 | 8.0 | 515 | 660 | [26] |
| $(Gd_{36}Tb_{20}Co_{20}Al_{24})_{97}Fe_{3}$ (MW, A+NC) | 108 | 5 | 7.6 | 520 | 680 | [26] |
| $Gd_{19}Tb_{19}Er_{18}Fe_{19}Al_{25}$ (MW, A+NC) | 97 | 5 | 5.94 | 569 | 733 | [24] |
| $Gd_{36}Tb_{20}Co_{20}Al_{24}$ (MW, A+NC) | 82 | 5 | 9 | 518 | 657 | [40] |
| $Dy_{36}Tb_{20}Co_{20}Al_{24}$ (MW, A+NC) | 42 | 5 | 8.2 | 301 | 414 | [40] |
| $Ho_{36}Tb_{20}Co_{20}Al_{24}$ (MW, A+NC) | 42 | 5 | 10.3 | 372 | 474 | [40] |
| $LaFe_{11.6}Si_{1.4}$ (MW, Amorphous) | 195 | 2 | 9.0 | - | 45 | [61] |
| $Ni_{48}Mn_{26}Ga_{19.5}Fe_{6.5}$ (MW, Amorphous) | 361 | 5 | 4.7 | - | 18 | [13] |
| $Gd_{53}Al_{24}Co_{20}Zr_{3}$ (B) | 93 | 5 | 9.4 | 509 | - | [49] |
| $Gd_{48}Al_{25}Co_{20}Zr_{3}Er_{4}$ (B) | 84 | 5 | 9.4 | 647 | - | [50] |
| $Gd_{55}Co_{20}Al_{25}$ (B) | 103 | 5 | 8.8 | 541 | - | [51] |
| $Gd_{55}Co_{25}Ni_{20}$ (B) | 78 | 5 | 8.0 | 640 | - | [51] |
| Gd (B) [b] | 294 | 5 | 10.2 | 410 | - | [4] |
| $Gd_5Si_2Ge_{1.9}Fe_{0.1}$ (B)[b] | 305 | 5 | 7.0 | 360 | - | [4] |

*SW: Single wire; MW: Multiple wires; B: Bulk; R: Ribbon; A+NC: Amorphous + Nanocrystalline;* [b] *Crystalline structure.*

## 5. Magnetocaloric microwires

While previous research focused on exploring the MCE in bulk materials (both single and polycrystalline forms) [1, 4, 5], the study of magnetocaloric materials in wire form has only drawn growing attention in the scientific community since the first work reported by F.X. Qin, et al. (2013) [12, 13] on the excellent mechanical, magnetic and magnetocaloric properties of melt-extracted $Gd_{53}Al_{24}Co_{20}Zr_3$ amorphous microwires. Since that work, a wide range of Gd-alloy microwires with outstanding magnetic and magnetocaloric properties has been fabricated using the melt-extraction technique [11-29]. Other types of magnetocaloric microwires such as Heusler and high entropy alloys $Ni_{48}Mn_{26}Ga_{19.5}Fe_{6.5}$ [13], Ni-Mn-Ga-Fe [30,33], Ni-Mn-Ga-Cu [31], Ni-Fe-Mn-Sn [34], Ni-Mn-Ga [32], HoErFe [36], DyHoCo [37], $RE_{36}Tb_{20}Co_{20}Al_{24}$ (*RE*= Gd, Dy or Ho), and $LaFe_{11.6}Si_{1.4}$ [61] have also been fabricated and investigated (Table 1).

### *5.1. Fabrication of microwires*

Figures 5A, B illustrate a fabrication method based on the melt-extraction technique advanced by Prof. Sun's group for making a variety of high quality Gd-alloy and Heusler alloy microwires [11-26, 30-33]. The whole fabrication process is under vacuum protection, so the chamber needs to reach a high vacuum of $10^{-3}$ Pa first and be then filled to $5\times10^3$ Pa with high purity Ar (99.97%). Before melting spin, an alloy rod is placed inside a borazon crucible with an inner diameter of 10.5 mm and the Cu wheel is pre-heated to 100°C prior to melt-extraction. The line velocity of the Cu wheel is typically fixed at 30 m/s, and the edge of the wheel shows an acute angle of 60° (Fig. 5A).

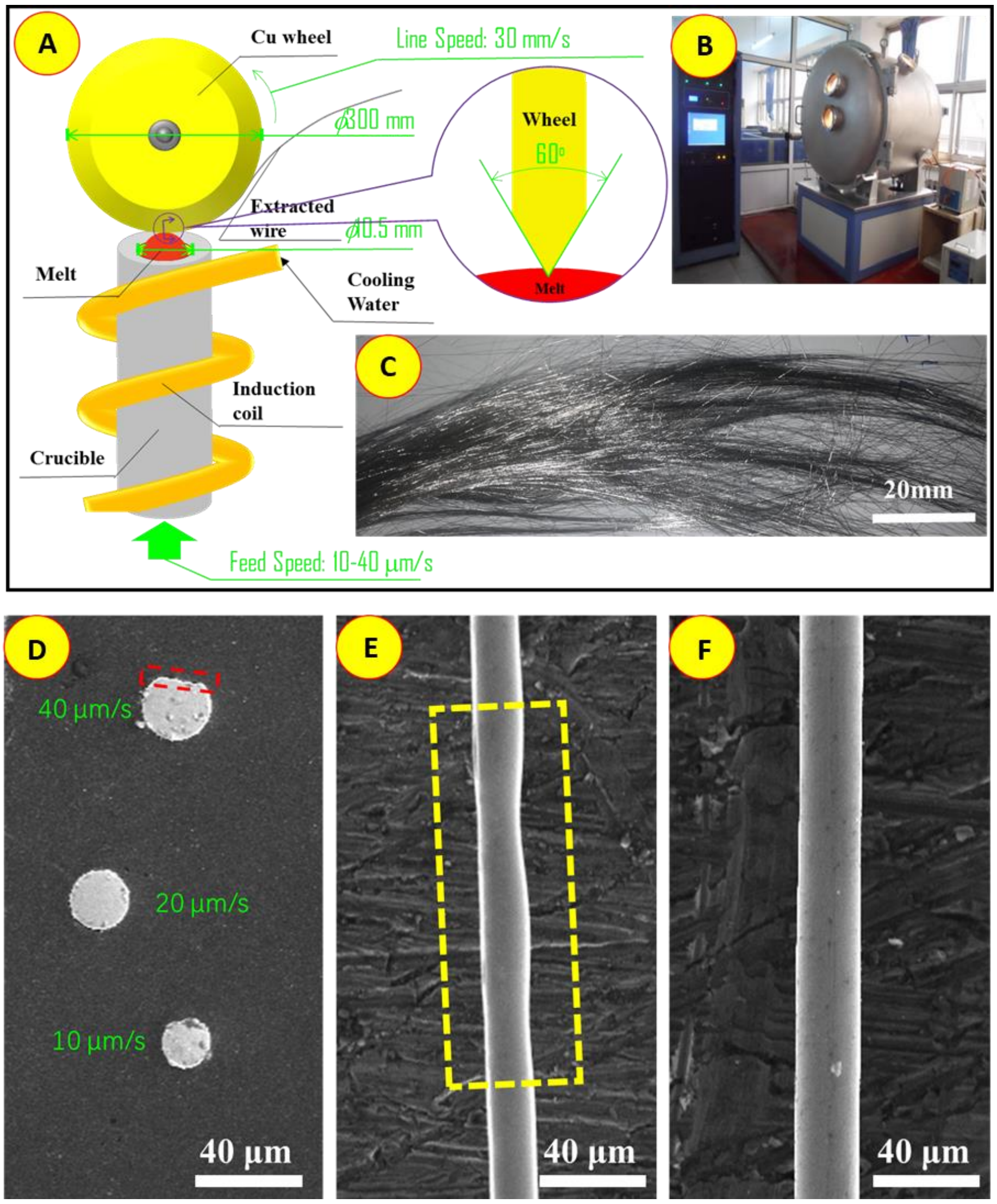

**Fig. 5. (A) Detailed illustration of the melt-extraction technique; (B) the melt-extraction equipment and (C) the macrophotograph of the fabricated wires. Cross-sectional (D) and SEM images of the Gd-based wires with different feed speeds of 10 μm/s (E) and 20 μm/s (F).**

The top of the rod is re-melt by high frequency electromagnetic induction heating, the melting temperature should be kept at $T_m$ + 50°C. To obtain desirable microwires, the feed speed can vary between 10 and 40 μm/s. When the melt is in contact with the edge of the wheel, a thin layer of the melt is extracted and adhered to the wheel. The melt layer is quenched to a supercooled state by the wheel and then deviated from the wheel due to the centrifugal force and decreasing wettability. Finally, the melt layer is rounded as the action of the surface tension and finally solidified to form long wires during free flight stage. The SEM images of Gd alloy melt-extracted microwires are shown in Fig. 5C. It is worth noting that an average diameter of wire can be varied by adjusting the feed speed (Fig. 5D-F). However, the uniformity of the wire diameter varies in a more complicated way. It is important to produce wires of uniform diameter over their long length.

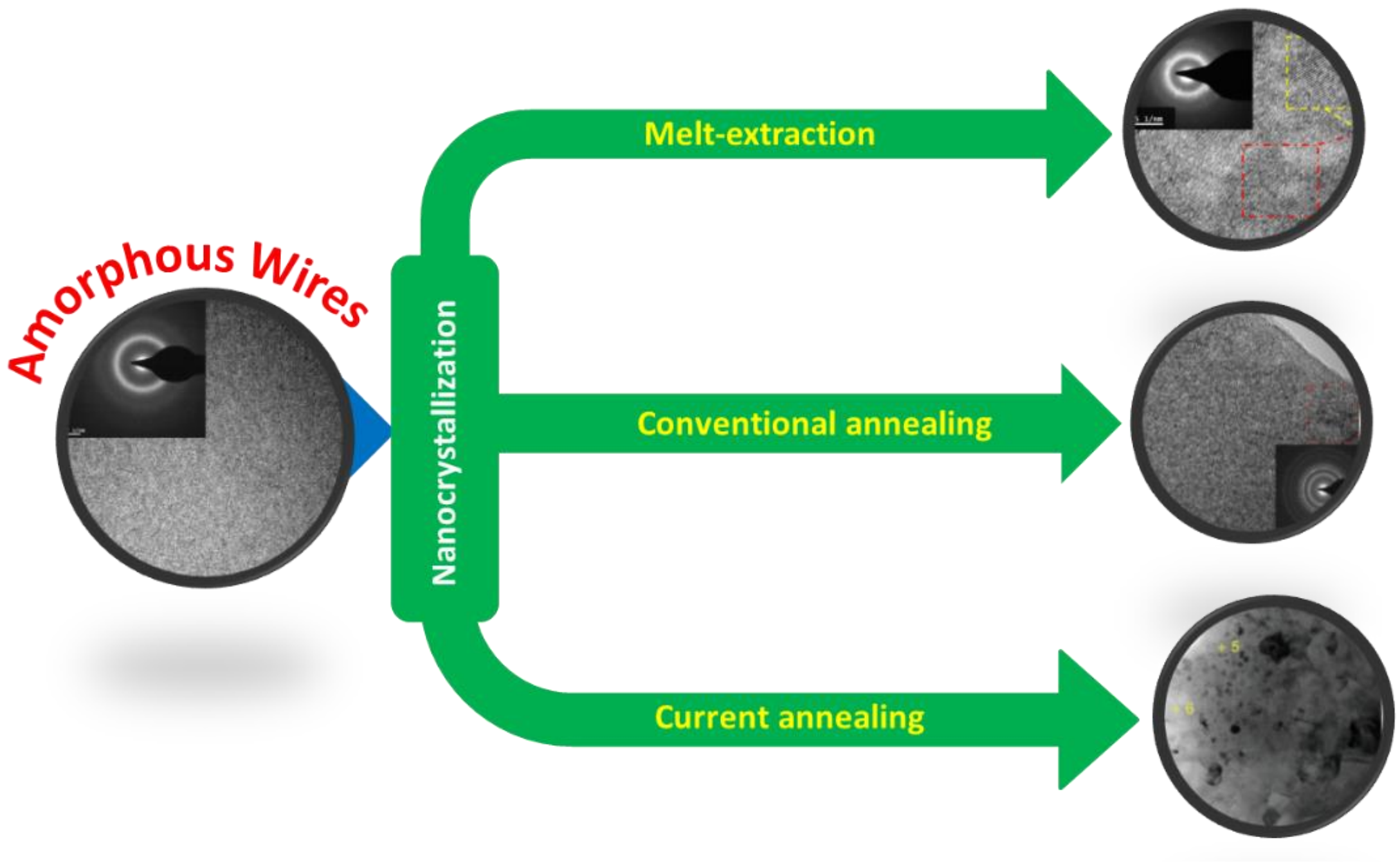

**Fig. 6. Three approaches for creating nanostructures in melt-extracted amorphous wires: nancrystallization via melt-extraction (approach #1); nancrystallization via conventional annealing (approach #2); and nancrystallization via current annealing (approach #3). High-resolution electron transmission microscopy (TEM) is often used to reveal the nature of nanostructures formed in melt-extracted wires.**

During melt-extraction, magnetocaloric microwires can be purely amorphous [11, 12, 20] or a mixture of amorphous and nanocrystalline phases [15, 16]. Adjusting the fabrication parameters associated with the melt-extraction process allows for the creation of either single phase (purely amorphous) wires or biphase amorphous/nanocrystalline wires with desirable magnetocaloric properties. Biphase amorphous/nanocrystalline wires can also be created by annealing their as-quenched amorphous counterparts [16]. The nanocrystalline to amorphous volume fraction ratio can be controlled by adjusting the annealing temperature and/or the annealing time. Based on these studies, we propose in Fig. 6 three approaches for creating nanostructures in melt-extracted magnetocaloric wires.

The first method is to utilize the advantage of controlling the parameters associated with melt-extraction. Nanocrystals can be created directly within an amorphous matrix of the wire during melt-extraction process [15]. The products are typically biphase nanocrystalline/amorphous wires. The shortcoming of this method is to sometime create unwanted minor phases in wires, which depend on alloy elements used. The second method is to create nanocrystals in amorphous wires through conventional (furnace) annealing [16]. This method also yields nanocrystalline/amorphous wires, in which nanocrystal sizes can be controlled by adjusting annealing temperature and time. This method is usually time-consuming since it takes time and effort to perform experiments to find out optimal annealing conditions. The third method is to utilize Joule heating effect by which the passage of a dc current through the wire produces heat that is sufficient to partially convert an amorphous phase (mostly on the surface of the wire) into the nanocrystalline phase [39]. We discuss below the advantages and shortcomings of these single-phase and biphase or multiphase magnetocaloric microwires. Table 1 lists several magnetocaloric microwires that are promising candidates for energy-efficient MR technology.

### *5.2. Gd-based microwires*

#### *5.2.1. Single phase materials*

Several single phase Gd-alloy microwires have been investigated, such as $Gd_{53}Al_{24}Co_{20}Zr_3$ [11, 12], $Gd_{50}(Co_{69.25}Fe_{4.25}Si_{13}B_{13.5})_{50}$ [19], $Gd_{60}Fe_{20}Al_{20}$ [20], and $Gd_{55}Co_{30}Ni_xAl_{15-x}$ [23] amorphous microwires. In addition to the superior properties that the wire form promises, their amorphous nature also brings additional advantages such as broadened operating temperature ranges and hence enhanced *RC* values due to their magnetically disordered structure.

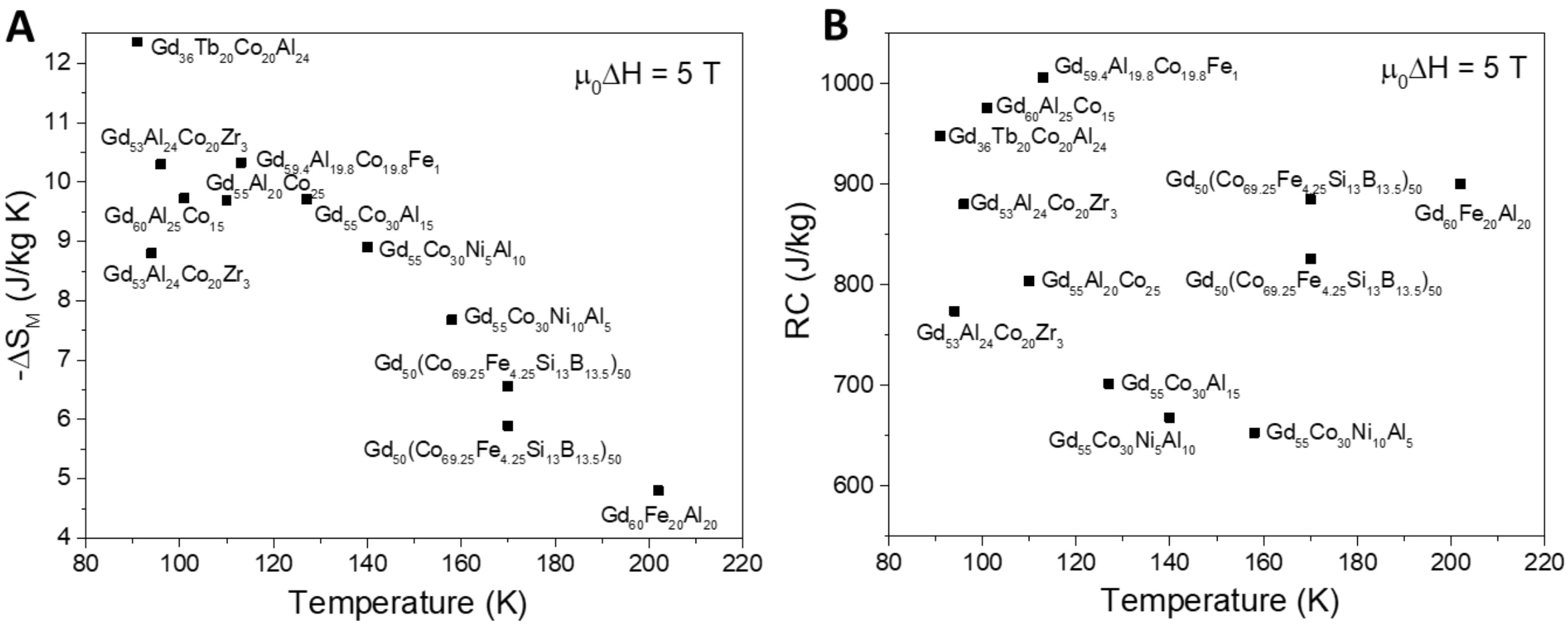

**Fig. 7. (A) Magnetic entropy change Δ$S_M$($T$) and (B) refrigerant capacity *RC*($T$) for a field change of 5 T for several single-phase Gd-alloy microwires.**

Figure 7 shows the Δ$S_M$ and *RC* values (at the corresponding Curie temperatures, $T_C$) of several single-phase Gd-based microwires for a field change of 5 T. It can be seen in this figure that both Δ$S_M$ and $T_C$ can be tuned by varying alloy compositions. There is a tendency that the larger values of Δ$S_M$ are achieved in materials with lower $T_C$ values. This tendency is rather different from that observed for biphase systems (Fig. 8). In the temperature range of 80-200 K, the $Gd_{36}Tb_{20}Co_{20}Al_{24}$ microwires show the largest Δ$S_M$ among samples compared (Fig. 7A), while the largest *RC* is achieved in the $Gd_{59.4}Al_{19.8}Co_{19.8}Fe_1$ microwires (Fig. 7B). The $Gd_{60}Fe_{20}Al_{20}$ microwires possess the largest *RC* around 200 K, due to the broadened Δ$S_M$($T$) distribution [20]. The *RC* values of the single-phase magnetocaloric microwires are also reported and compared in Fig. 9 along with the biphase magnetocaloric microwires.

*5.2.2. Biphase materials*

While single phase magnetocaloric microwires have been shown to exhibit superior magnetocaloric properties, there is an opportunity for further improvement in their magnetocaloric figures of merit. This can be achieved by creating magnetic nanocrystals embedded in an amorphous magnetic matrix [15, 16]. Magnetic coupling between the crystalline and amorphous phases can be tuned to achieve optimal magnetic and magnetocaloric properties, while preserving the good mechanical responses of the microwires [16].

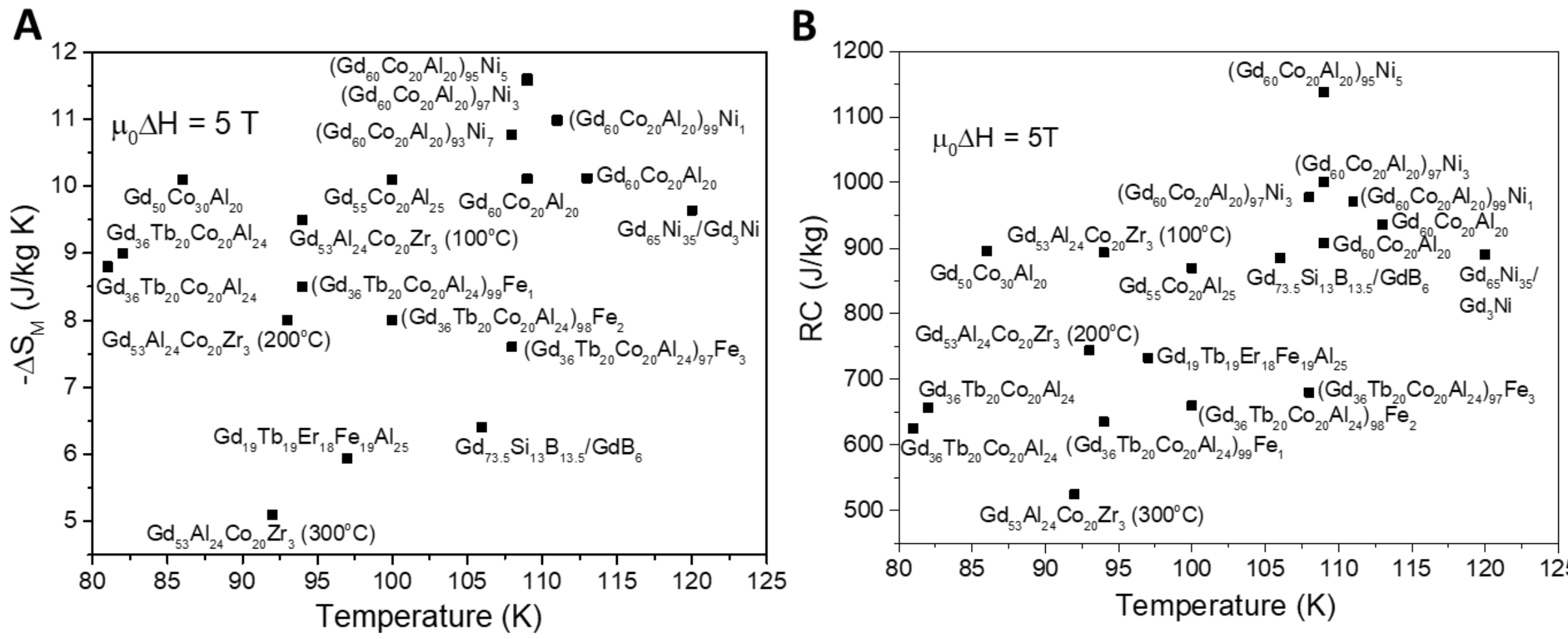

**Fig. 8. (A) Magnetic entropy change Δ*S*M(*T*) and (B) refrigerant capacity *RC*(*T*) for a field change of 5 T for several biphase Gd-alloy microwires.**

The first method has been proposed (Fig. 6) and employed to create a novel biphase amorphous/nanocrystalline structures in Gd-based microwires by controlling the melt-extraction process [15, 21, 22]. Studies have shown that a large RC enhancement in biphase amorphous/nanocrystalline Gd-Al-Co wires can be obtained directly from the adjusting the melt-extraction process, relative to their amorphous counterparts [15]. The same method has been applied to create $Gd_{73.5}Si_{13}B_{13.5}/GdB_6$ [21] and $Gd_3Ni/Gd_{65}Ni_{35}$ ferromagnetic/antiferromagnetic microwires [22]. Fig. 8 shows the $\Delta S_M$ vs. $T_C$ plot for several biphase Gd-based microwires at $\mu_o H$ = 5 T. For a field change of 5 T, $\Delta S_M$ was ~9.64 J/kg.K at ~120 K for the $Gd_3Ni/Gd_{65}Ni_{35}$ microwires, which is 1.4 times larger than that reported for $Gd_{65}Ni_{35}$ ribbons [22]. In the temperature range of 80-120 K, the biphase nanocrystalline/amorphous $(Gd_{60}Co_{20}Al_{20})_{95}Ni_5$ microwires show the largest values of both $\Delta S_M$ and *RC* among samples compared (Fig. 8A, B). It can be seen in Fig. 8 that the optimal biphase microwires created directly from melt-extraction tend to show enhanced values of $\Delta S_M^{max}$ and *RC* as compared to their single phase components. However, it is not straight forward to find out what phases can be possibly formed during melt-extraction.

The second method has been employed through thermal annealing of as-prepared amorphous microwires (Fig. 6) [16]. The annealing temperature was adjusted to achieve optimization of nanocrystals embedded in an amorphous matrix in $Gd_{53}Al_{24}Co_{20}Zr_3$ microwires. These biphase microwires were created by annealing as-quenched ones at 100, 200, and 300°C [16]. A detailed analysis of the magnetic and magnetocaloric properties of the annealed microwires revealed the optimal annealing conditions. Some small islands of nanocrystallites with an average size of 5, 8, and 10 nm were uniformly distributed in the amorphous matrix for the $Gd_{53}Al_{24}Co_{20}Zr_3$ microwires treated at 100, 200, and 300°C, respectively. It was found that the sample annealed at 100°C achieved the optimal $\Delta S_M^{max}$ and *RC* values of 9.5 J/kg.K and 687 J/K. This indicates that the appropriate annealing of as-prepared $Gd_{53}Al_{24}Co_{20}Zr_3$ amorphous microwires could enhance both $\Delta S_M^{max}$ and *RC* significantly. The mechanical properties of the microwires were also slightly improved upon utilizing these annealing conditions. This could be attributed to the presence of small nanocrystals and their uniform distribution in an amorphous matrix. By contrast, annealing at higher temperatures (200 and 300$^o$C) was found to reduce the $\Delta S_M^{max}$ and *RC* considerably (Table 1). The mechanical properties of the microwires were also degraded. This could be associsted with an increase in size of magnetic nanocrystals and their less uniform distribution in the amorphous matrix. The main drawback of this method is that microwires must be annealed over a relatively wide range of annealing temperatures and structurally and magnetically characterized, in order to determine at which annealing temperature an optimal size of nanocrystals and their uniform distribution can be achieved.

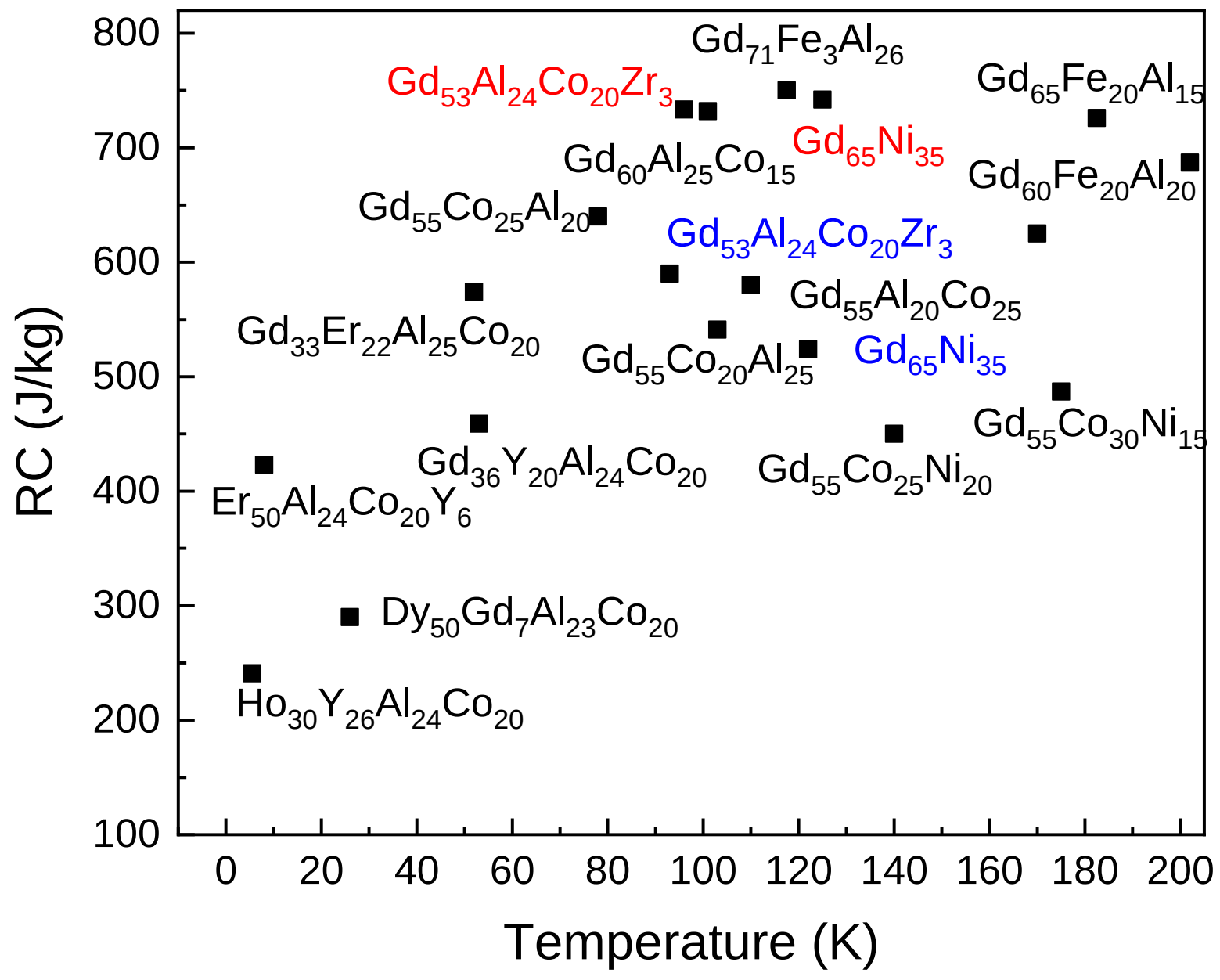

**Fig. 9.** ***RC*** **values of candidate materials in the temperature range of 10-200 K. Red indicates the microwires, while blue indicates their bulk counterparts for comparison.**

For comparison, the *RC* values of several Gd-based microwires and other materials for a field change of 5 T are plotted in Fig. 9. Clearly, the $Gd_3Ni/Gd_{65}Ni_{35}$ and $Gd_{53}Al_{24}Co_{20}Zr_3$ microwires possess the largest *RC* values among the candidates compared. As compared to Gd and $Gd_5Si_2Ge_{1.9}Fe_{0.1}$ alloys, these Gd-alloy microwires possess the larger MCEs and *RC*s (Table 1), although their Curie temperatures are much lower (below 200 K). The *RC* values of the Gd-alloy microwires are also greater than those reported for their bulk and ribbon counterparts (see Table 1). The excellent magnetocaloric and mechanical properties of the Gd-based magnetocaloric microwires make them very promising candidates for active MR in the liquid nitrogen temperature regime.

### *5.3. Heusler alloy microwires*

Although the above-discussed Gd-alloy microwires with excellent magnetic and magnetocaloric properties are promising candidates for MR, their Curie temperatures are limited to temperatures below 80-200 K (Figs. 7 and 8), reducing their cooling applications at ambient temperature. It is therefore essential to develop new types of microwires that exhibit large MCEs and *RC*s in the room temperature region.

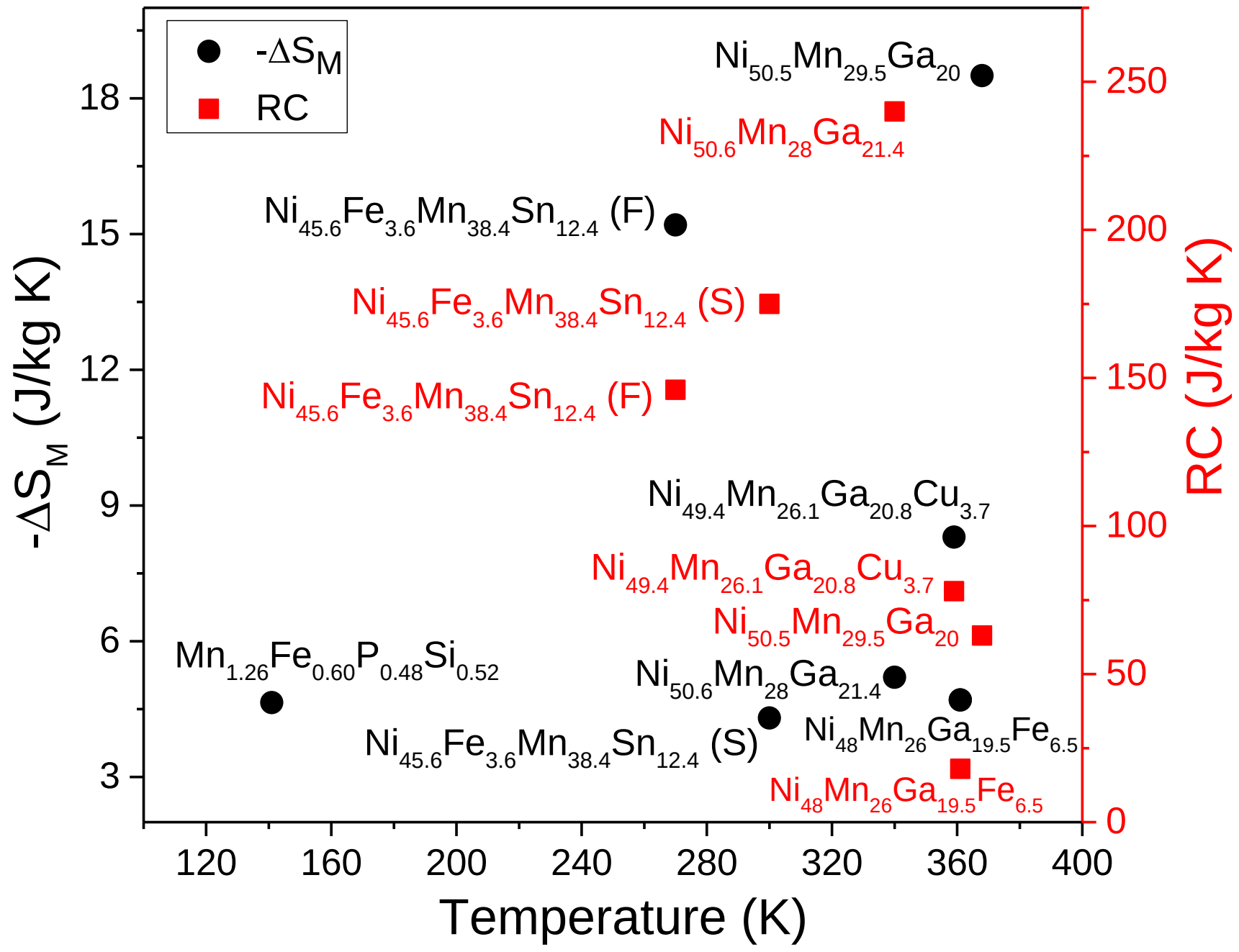

**Fig. 10. Magnetic entropy change Δ*S*M(*T*) and refrigerant capacity *RC*(*T*) for a field change of 5 T for several Heusler alloy microwires.**

Heusler alloys, such as Ni-Mn-Ga, are known to undergo magneto-structural phase transitions (also known as martensitic transitions) and exhibit the large MCEs around these transitions [30-35]. It is worth mentioning that the magnetization and its change associated with the martensitic transition in these Heusler alloys strongly depend on the valence electron concentration per atom e/a, which can be modulated by chemical doping Fe, Co, Cu, In, Ge, etc [52, 53]. As a result, the magnetic and magnetocaloric properties of the Heusler alloys can be tuned over a wide temperature range. Fig. 10 shows the $\Delta S_M$ and *RC* values of several Heusler alloy microwires for a field change of 5 T. Among the systems compared, the $Ni_{50.5}Mn_{29.5}Ga_{20}$ microwires possess the largest $\Delta S_M$, while the largest *RC* is achieved in the $Ni_{50.6}Mn_{28}Ga_{21.4}$ microwires. Interestingly, these large values are achieved at ambient temperature. Table 1 also compares the magnetocaloric properties of the Heusler alloy microwires with other candidate materials. It is worth noting that most of the Heusler alloy microwires order ferromagnetically (e.g., the SOMT) at or above room temperature, followed by the martensite phase transition (e.g., the FOMT) at lower temperatures [30-35]. The larger MCEs are often observed around the FOMT but concentrated on a narrower temperature region. By contrast, smaller MCEs are found around the SOMT but span over a much larger temperature range. As a result, the *RC*s are found to be even larger around the SOMT in some Heusler alloy microwire systems. It has also been reported that large hysteretic losses occur in these systems, associated with the FOMT nature which significantly reduces the RC of the material [30-35]. By refining the chemical compositions in these systems, it is possible to reduce the hysteretic losses, resulting in improvement of the *RC*s while preserving their large MCEs.

### *5.4. High-entropy alloy microwires*

Recently, there has been a growing interest in developing high entropy magnetocaloric materials for MR, due to their excellent mechanical properties [54]. A number of high-entropy microwire systems with good magnetocaloric responses have been reported [37, 41]. Interestingly, H. Yin, et al. (2022) [39] reported that the magnetocaloric properties of high entropy alloy $(Gd_{36}Tb_{20}Co_{20}Al_{24})_{100-x}Fe_x$ microwires can be enhanced and optimized by current annealing of their as-cast amorphous microwire counterparts. The precipitation of nanocrystals within an amorphous matrix creates a phase compositional difference in the microwires, resulting in a broadened MCE and hence an enhanced *RC* in the current-annealed microwires. In addition to fine tuning chemical compositions, this current-annealing method provides an additional degree of control over microstructure in high-entropy alloy microwire systems (Fig. 6), allowing for MCE and *RC* tunability. Table 1 also lists the magnetocaloric properties of several high-entropy alloy microwires for MR in the low and high temperature regions. It seems that the high-entropy alloy microwires tend to show reduced MCEs over larger temperature ranges compared to other magnetocaloric candidate materials. Conventional/current annealing of the amorphous microwire counterparts may lead to enhancement of the MCE and *RC*, but it is also important to preserve their excellent mechanical responses.

## 6. Design of magnetic beds composed of table-like magnetocaloric multiwire arrays

For magnetic cooling applications above 20 K, the regenerative Ericsson cycle is theoretically ideal for magnetic cooling systems due to its high working efficiency and broad temperature range [4]. Thus, magnetic refrigerants showing broad working temperature ranges are required which means the materials should have a table-like MCE near the $\Delta S_M^{max}$ [55-59]. Typically, the shapes of $\Delta S_M(T)$ for alloys or elementary substances are shape peaks or broad triangular peaks [1, 4]. Materials with different phases usually show different MCE properties such as different Curie temperatures and the MCEs of these alloys are manifested as combinations of these respective phases [55-59]. The multiphases are commonly formed during the fabrication process [55] or obtained through thermal annealing [57]. It is quite challenging to control the amount of each individual phase or obtain the required phase. A multiphase composite can also be designed by selecting alloys with different compositions and MCE values [59].

In this context, magnetocaloric microwires possess superior advantages as they can be easily assembled into multiwire arrays or laminate structures. To design a multiwire-based magnetic bed that exhibits table-like magnetocaloric characteristics, it is essential to select microwires with similar magnetocaloric properties and close Curie temperatures [27-29]. To be simple, the wires can be mixed mechanically and assumed without any interactions among themselves, the magnetic entropy changes of the multiwire bed can be calculated as [29]

$$\Delta S_M^{Design} = \alpha \Delta S_M^1 + \beta \Delta S_M^2 + \gamma \Delta S_M^3 + \cdots\cdots + \omega \Delta S_M^n \tag{6}$$

where $\Delta S_M^{Design}$ is the magnetic entropy change of designed structures, $\alpha$, $\beta$, $\gamma$ and $\Delta S_M^1$, $\Delta S_M^2$, $\Delta S_M^3$ are the weight fractions and magnetic entropy changes of the different microwires.

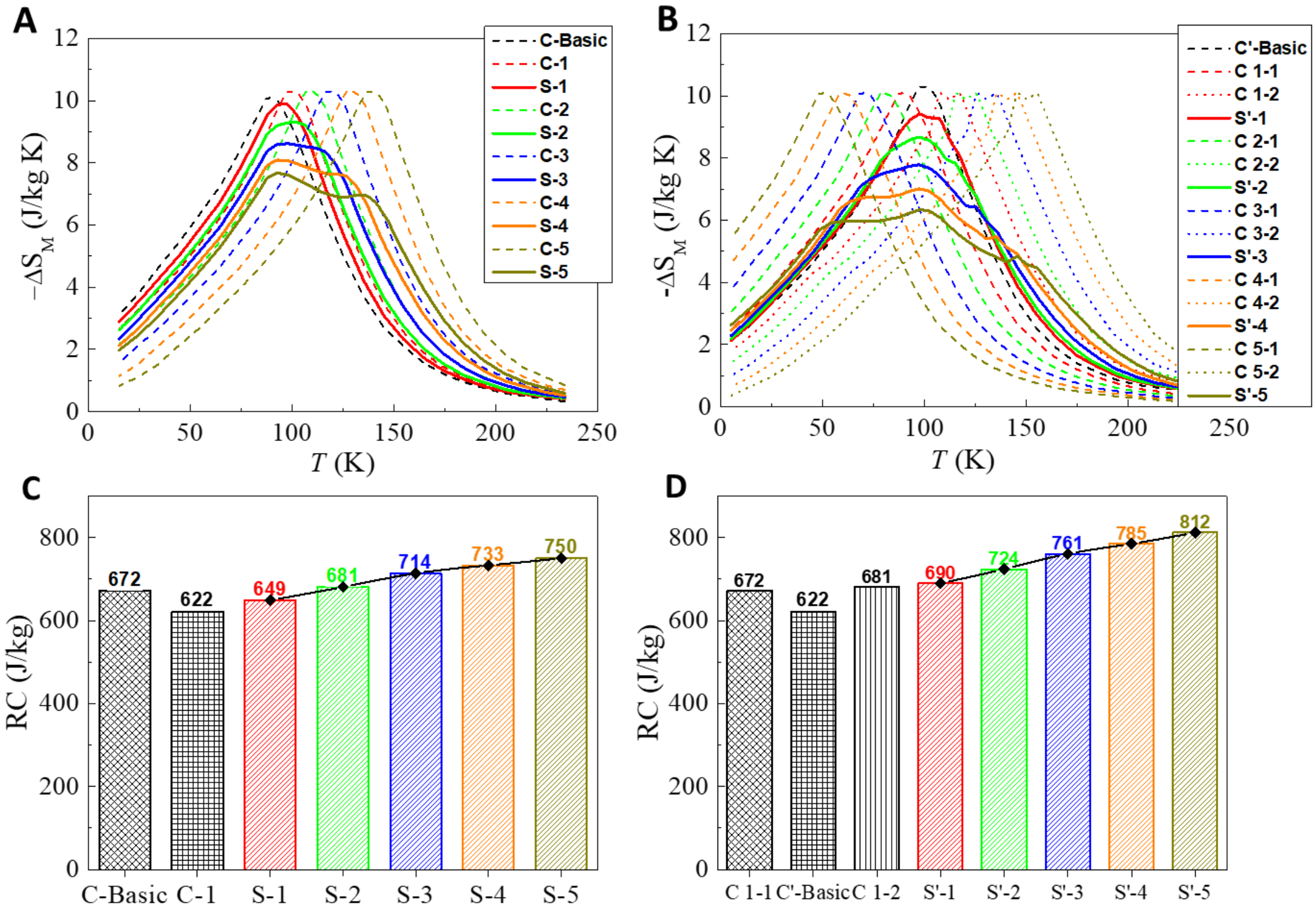

**Fig. 11. Simulated -Δ$S_M$ *vs.* *T* curves at $\mu_0\Delta H$=5 T of the designed systems and their components for (A) two-component systems and (B) three-component systems. *RC* values of (C) Sample C-Basic and (D) Sample C'-Basic and their components. Adapted from Ref. [29].**

By selecting Δ$S_M$ and *RC* values for the designed structures with different $T_C$ intervals, it is possible to design different samples named S-1, S-2, S-3, S-4, and S-5 by two components with different $T_C$ intervals. Herein, the Δ$S_M$(T) data of $Gd_{50}Al_{30}Co_{20}$ microwires was used as one basic composition (named C-Basic) and $Gd_{50}Al_{25}Co_{25}$ microwires as other varying compositions (C-1, C-2, C-3, C-4 and C-5). The $T_C$ intervals between varying compositions (C-1, C-2, C-3, C-4, C-5) and C-Basic are ~10 K, 20 K, 30 K, 40 K and 50 K, respectively. The -Δ$S_M$ *vs.* *T* curves at $\mu_0\Delta H$=5 T for the designed structures are calculated using Eq. (6) are displayed in Figs. 11A, B, which shows that the designed samples (S-1, S-2, S-3, S-4, and S-5) exhibit more table-like MCE with the increasing $T_C$ intervals. Additionally, the values of -Δ$S_M^{max}$ decreases when the $T_C$ intervals increase. The RC values of the designed samples and their components (C-1, C-2, C-3, C-4, and C-5) at $\mu_0\Delta H$=5 T are plotted in Figs. 11C, D for comparison. The *RC* value of S-1 is lower than that of C-Basic when $T_C$ interval is ~10 K, while RC values of other samples are higher than those of both components. Remarkably, the *RC* values of the designed samples increase with increasing the $T_C$ interval, which means that the enhanced *RC* values for the designed systems can be obtained by using the Gd-based wires with large $T_C$ intervals in a range of temperature. The designed systems with three components show the same tendency as those with two components, which exhibits a table-like MCE and the values of $-\Delta S_M^{max}$ decreases when the $T_C$ intervals increase (Fig. 11D).

**7. Concluding remarks and outlook**

Refrigerants in form of microwires can yield higher cooling efficiencies compared to their bulk or other shape counterparts. The use of these microwires promises high applicability in active MR due to their outstanding soft magnetic, magnetocaloric and mechanical properties. The micro-size of the microwires enhances the surface area for the designed structures, which improves the efficiency of heat transfer significantly in cooling systems. These excellent magnetocaloric and mechanical properties can be harnessed by creating novel biphase nanocrystalline/amorphous structures via thermal annealing or during melt-extraction. These functional magnetocaloric microwires can be assembled as laminate structures, rendering them applicable in active cooling devices for MEMS and NEMS.

A variety of magnetocaloric microwires have been designed, fabricated, and investigated (Table 1). It has been shown that the Gd-alloy microwires are promising candidates for active MR in the liquid nitrogen temperature regime. Due to their SOMT, these microwires show no hysteresis losses, along with large MCEs and RCs. While alloying or doping has been found to tune the Curie temperature of the Gd-alloy microwires between 80 and 200 K, it appears that achieving these superior properties at ambient temperature is challenging in Gd-based microwire systems. Heusler alloy microwires such as Ni-Mn-Ga [32], Ni-Mn-Ga-Fe [30,33], Ni-Mn-Ga-Cu [31], and Ni-Fe-Mn-Sn [34] are interesting candidates for cooling applications in the room temperature regime, since they order ferromagnetically at or above room temperature. While these systems show large MCEs around the FOMT transition, the *RC*s are found to be much smaller compared to those around the SOMT transition, the MCEs of the latter are much smaller though. The disadvantage of these microwires also lies in the large hysteretic losses associated with the FOMT nature. Refining chemical compositions/nanostructures in these systems may offer an approach for reducing the hysteretic losses and improving the *RC*s while preserving the large MCEs [60]. High entropy alloy microwires are also perspective candidates for cooling applications, but their MCEs and *RC*s are relatively smaller compared to the Gd-alloy and Heusler alloy microwires.

Moving forward, further research should be undertaken to examine the applicability of the currently available magnetocaloric microwires in cooling devices. It is essential to quantify how high the cooling energy efficiency can be achieved in a cooler using the microwires compared to their other shape counterparts. Several challenges need to be overcome. For instance, it is important to improve the existing melt-extraction techniques to create magnetocaloric microwires with uniform diameters. It is also challenging to control the fabrication parameters associated with the melt-extraction process for achieving pure amorphous phase microwires or biphase microwires with controlled amorphous/nanocrystalline volume fraction ratios. The design of magnetocaloric microwires with controlled large, broadened magnetic entropy changes around room temperature is a very important but challenging task. The *RC*s of these microwires can be enhanced by coating the microwires with soft ferromagnetic layers such as FeNi, creating core/shell wire structures. Interfacial magnetic coupling and interaction between the core and shell is expected to broaden the magnetic entropy change, giving rise to the RC enhancement. Combing the melt-extraction and electrodeposition methods is considered most appropriate for making these novel core/shell wire systems. In an array of magnetic microwires, magnetostatic interactions can play an important role in determining the magnetocaloric response of the system. In this context, tuning magnetostatic interactions through incorporating short magnetocaloric microwires into a thermally conductive polymer matrix appears to be an interesting prospective approach for creating novel magnetic refrigerants. Combined theoretical and experimental studies are needed to validate these hypotheses. Bringing the fundamental studies to the point of making prototype refrigeration devices will require the concerted effort of interdisciplinary scientists and engineers.

Since the Heusler alloy microwires [30-34] exhibit ferromagnetic order above room temperature, they may also be promising candidates for magnetic/temperature sensing [62, 63], magnetic hyperthermia [64], and electromagnetic microwave applications [65, 66]. Harnessing these novel functionalities might make them truly multifunctional for modern device applications, opening new opportunities for research in this interdisciplinary area.

## COMPETING INTERESTS

The authors declare that there is no conflict of interest regarding the publication of this article.

## CRediT author statement

H.X. Shen, N.T.M. Duc, H. Belliveau, L. Luo, and Y.F. Wang performed experiments and analyzed the data. H.X. Shen, N.T.M. Duc, H. Belliveau, and M.H. Phan wrote the initial manuscript with inputs from J.F. Sun and F.X. Qin. M.H. Phan led the project.

## ACKNOWLEDGEMENTS

This work was supported the National Natural Science Foundation of China (NSFC, Nos. 51801044, 51671071, 51871124, 51701099 and 51561026).